\documentclass[aps,pre,twocolumn,superscriptaddress,showpacs]{revtex4}
\usepackage{epsfig}

\newcommand{\rem}[1]{\textit{\tiny #1}}

\begin{document}

\title{Aging in attraction-driven colloidal glasses}
\author{Antonio M. Puertas}
\affiliation{Group of Complex Fluids Physics, Department of Applied Physics, University of Almeria, 04120 Almeria, Spain}
\author{Matthias Fuchs}
\affiliation{Fachbereich Physik, University of Konstanz, D-78457 Konstanz, Germany}
\author{Michael E. Cates}
\affiliation{SUPA, School of Physics, The University of Edinburgh, JCMB Kings Buildings, Mayfield Road, Edinburgh EH9 3JZ, UK}
\date{\today}

\begin{abstract}
Aging in an attraction-driven colloidal glass is studied by computer simulations. The system is equilibrated without attraction and instantaneously ``quenched'', at constant colloid volume fraction, to one of two states beyond the glass transition; one is close to the transition, and the other one deep in the glass. The evolution of structural properties shows that bonds form in the system, increasing the local density, creating density deficits (holes) elsewhere. This process slows down with the time elapsed since the quench. As a consequence of bond formation, there is a slowing down of the dynamics, as measured by the mean squared displacement and the density, bond, and environment correlation functions. The density correlations can be time-rescaled to collapse their long time (structural) decay. The time scale for structural relaxation shows for both quenches a super-linear dependence on waiting time; it grows faster than the bond lifetime, showing the collective origin of the transition. At long waiting times and high attraction strength, we observe {\rem completely} arrested dynamics for more than three decades in time, although individual bonds are not permanent on this time scale. The localization length decreases as the state moves deeper in the glass; the non-ergodicity parameter oscillates in phase with the structure factor. Our main results are obtained for systems with a barrier in the pair 
potential that inhibits phase separation. However, when this barrier is removed for the case of a deep quench, we find changes in the static structure but almost none in the dynamics. Hence our results for the aging behavior remain relevant to experiments in which the glass transition competes with phase separation.
\end{abstract}

\maketitle

\newpage

\section{Introduction}

When a liquid is quenched below its glass transition temperature, its dynamics slows down enormously and the structural relaxation time becomes bigger than the experimental time: an amorphous solid is formed. In structural glasses, in contrast to spin glasses where theoretical approaches have been developed \cite{cugliandolo}, no theory is yet available to describe the slowing down of states beyond the glass transition, {\em i.e.}, aging. Computer simulations have provided, nevertheless, important information about general properties of aging \cite{kob97}.

In colloidal systems with short range attractions, vitrification can be repulsion driven (induced by increasing the pressure) or attraction driven (induced by decreasing temperature at intermediate densities) \cite{pham02,sciortino02,dawson02}. The former is caused by the steric hindrance of hard cores at high densities (as is the glass transition in most atomic or molecular systems). The latter is more specific to colloids with short range attractions. It is induced by increasing the attraction strength, and caused by bonding between the particles. At high attraction strength and high density, where both glass transitions merge, a higher order singularity appears, causing a logarithmic decay of the density correlation function \cite{puertas02,sciortino03,pham04,sperl04}. Mode coupling theory (MCT) anticipated this scenario\cite{bergenholtz99,fabbian99,dawson01}, and correctly describes the liquid states close to both glass transitions, where the structural slowing down sets in \cite{puertas02,foffi02,zaccarelli02,pham04,puertas05}. However, it does not provide information about aging within the arrested state.

Using molecular simulations, Kob {\em et al} studied the aging of a structural (repulsive) glass of Lennard-Jones particles \cite{kob97}. Quenching from high temperatures, the slowing down of the dynamics was studied by means of a two-time density correlation function, $\Phi_q(t,t_w)$, where the earlier of the two times, $t_w$, represents a `waiting time' after the quench. They found a scaling of the structural relaxation (similar to the time--temperature scaling in fluid states) with the relaxation time following a power law with $t_w$. More recent studies within the nonergodic region have focused on dynamical heterogeneities and their importance to the glass transition \cite{doliwa02,vollmayr02,vollmayr05}.

Recent simulations of the attraction driven glass have considered systems where structural arrest interplays with liquid-gas separation, producing states of arrested phase separation \cite{gimel02,darjuzon03,foffi05}.  The observed phenomena from simulations resemble the experimental observations \cite{kroy04}, but the dynamics of the system was not analyzed in terms of glass aging. In this paper we isolate the effects of the attractive glass transition from those of phase separation, and study the resulting aging dynamics without this additional complication. One paper has addressed the attractive glass by studying a system at high concentrations, above the onset of the single-phase liquid \cite{foffi04}. Starting from the fluid pocket between the attractive and repulsive glasses, the repulsive glass could be reached by increasing the temperature, and the attractive glass by decreasing it. However, the states studied are close to the higher order singularity mentioned above; this affects the dynamics near the glass transition \cite{sciortino03}, and possibly also its aging within the non-ergodic regions. A different work by the same group \cite{zaccarelli03} studied the transition from repulsive glass to attractive glass at high density, and it was concluded that the bonds were not strong enough stabilize the attractive glass and the glass to glass transition was not observed. Introducing an artificial barrier to stabilize these bonds, arrested dynamics due to the attractive glass was finally obtained and the glass to glass transition observed \cite{zaccarelli03}.

In this work, we study a system which we have used previously to analyze the fluid close to the attraction driven glass transition \cite{puertas02,puertas05,puertas03}. The liquid-gas transition has been inhibited by introducing a long-range repulsive barrier in the interaction potential \cite{puertas02}. We study the aging of the system in two different states, on an isochore where we have previously determined the transition point. On performing an instantaneous quench from the system without attraction, the structure and dynamics of the system are monitored. As bonds between particles form, the local density increases, causing holes to open up elsewhere in the system. Accordingly, structural relaxation takes longer and longer as aging proceeds.

We show below that the attractive glass is broadly similar to the repulsive one (and thus to atomic glasses) in many fundamental properties, such as the time-scaling of correlation functions and the evolution of the localization length. An important difference is also noticed between the two states studied here: the glass close to the transition is ``softer'' than that deeper in the non-ergodic region, where the dynamics indeed arrests. The bond correlation function shows that the bonds are reversible even for the strongest attraction studied. The bond and density correlation functions evolve differently, in contradiction with recent results found in a different system at high temperatures and by a different simulation method \cite{foffi05b}. Finally, we turn off the repulsive barrier and study the competition between the attractive glass transition and liquid-gas separation, to elucidate the role of the barrier in the dynamics. This investigation is motivated in part by recent observations of equilibrium clusters in protein solutions at lower packing fractions where a charge induced repulsive barrier is considered essential \cite{stradner04}. The results show that the dynamics of the system is similar with and without the barrier. This suggests that the aging dynamics is primarily controlled by bond formation in a state of arrested phase separation, rather than slow evolution of the long-wavelength density fluctuations that are present in such a state. A companion study of dynamic heterogeneities in the attractive glass will be presented elsewhere \cite{puertas06}.

\section{Simulation details}

The simulated system consisted of 1000 quasi-hard particles interacting by a short range attraction, with Newtonian dynamics.  The system is polydisperse to avoid crystallization: particle sizes are distributed according to a flat distribution of half--width $\delta=0.1 a$, where $a$ is the mean radius. The core-core repulsion is given by

\begin{equation}
V_{sc}(r)\:=\:k_BT \left(\frac{r}{2a_{12}}\right)^{-36}
\end{equation}

\noindent where $2a_{12}=(a_1+a_2)$, with $a_1$ and $a_2$ the radii of the interacting particles. The short range attraction mimics the interaction between colloidal particles induced by non-adsorbing polymers in a colloid polymer mixture. For a polydisperse colloidal system, the interaction is \cite{mendez00}:

\[ V_{AO}(r) \:=\: -k_BT \phi_p \left\{\left[\left(\bar{\eta}+1\right)^3 -\frac{3r}{4\xi} \left(\bar{\eta}+1\right)^2+\frac{r^3}{16\xi^3}\right]+
\right.\]
\begin{equation}\label{pot} 
\left.+\frac{3\xi}{4r} \left(\eta_1-\eta_2\right)^2 \left[\left(\bar{\eta}+1\right) -\frac{r}{2\xi} \right]^2\right\}
\end{equation}

\noindent for $2a_{12} \leq r \leq 2(a_{12}+\xi)$ and $0$ for larger distances. Here, $\eta_i=a_i/\xi$; $\bar{\eta}=(\eta_1+\eta_2)/2$, and $\phi_p$ is the volume fraction of the polymer. The range of the interaction, $\xi$, is the polymer size, and its strength is proportional to $\phi_p$, the concentration of ideal polymers. This potential has been slightly corrected near $r=2a_{12}$ to ensure \cite{puertas03} that the total interaction potential has its minimum at $2a_{12}$.

At high polymer fractions, or attraction strengths, this system would undergo liquid-gas separation into two fluid phases. The interplay of this demixing with the attractive glass transition is complex \cite{kroy04}, and would complicate the interpretation of our results. To avoid it, we add a repulsive long-range barrier to the interaction potential:

\begin{equation}
V_{bar}(r)\:=\:k_BT\left\{\left(\frac{r-r_1}{r_0-r_1}\right)^4-2\left( \frac{r-r_1}{r_0-r_1}\right)^2+1\right\}
\end{equation}

\noindent for $r_0\leq r \leq r_2$ and zero otherwise, with $r_1=(r_2+r_0)/2$. The limits of the barrier were set to $r_0=2(a_{12}+\xi)$, and $r_2=4a$, and its height is only $1 k_BT$ (equal to the attraction strength at $\phi_p=0.0625$). The barrier raises the energy of a dense phase, so that liquid-gas separation is suppressed. The resulting total interaction, $V_{tot}=V_{sc}+V_{AO}+V_{bar}$, is analytical everywhere and allows straightforward integration of the equations of motion. A short range attraction complemented by long range repulsive may induce microphase separation \cite{sear99,imperio04}. We have performed some simulations without the repulsive barrier to test its effects on the dynamics of the system, as shown in Sec. III.C.

Lengths are measured in units of the average radius, $a$; time in units of $\sqrt{4a^2/3v^2}$, where the thermal velocity $v$ is set to $\sqrt{4/3}$. Equations of motion are integrated using the velocity-Verlet algorithm, in the canonical ensemble (constant NTV), to mimic the colloidal dynamics. Every $n_t$ time steps, the velocity of the particles is re-scaled to assure constant temperature. The time step was set to $0.0025$ and $n_t=100$. The range of the attraction, $2\xi$, is set to $2\xi=0.2$. The density of colloids is reported as volume fraction, $\phi_c=\frac{4}{3}\pi a^3 \left(1+\left(\frac{\delta}{a}\right)^2\right) n_c$, with $n_c$ the colloid number density; the attraction strength is fixed by the polymer volume fraction, $\phi_p$. (Note that, with $\xi=0.1$, the attractive minimum is, for average sized particles, at $r=2a$, with $V_{min}=-16 k_BT \phi_p$.) 

In our simulations the system is first equilibrated at $\phi_p=0$ and $\phi_c=0.40$ (infinite temperature) with the repulsive barrier, and then instantaneously ``quenched'', at constant colloid density, to the desired attraction strength. The time elapsed since the quench is called the {\em waiting time}, $t_w$. Both the static and dynamic quantities depend on $t_w$; {\em i.e.}, structural properties become waiting-time dependent, and dynamical ones are two-time functions, depending on  $t_w$ (the first of the two times) and $t'=t-t_w$, with $t$ the second time and $t'$ the interval. For instance, the density correlation function reads,

\begin{equation}
\Phi_q(t',t_w)\:=\:\frac{1}{N^2} \sum_{j,k} \exp\left\{i {\bf q}\left({\bf r}_j(t'+t_w)-{\bf r}_k(t_w)\right)\right\}
\end{equation}

\noindent where $j$ and $k$ run over all $N$ particles in the system. The mean squared displacement, the self part of the density correlation function, or bond correlation functions, can be defined similarly.

The system is quenched to two states inside the non-ergodic region, $\phi_p=0.50$ and $\phi_p=0.80$ (both with colloid density $\phi_c=0.40$). For each state, 50 independent simulations were run, to improve the ensemble average. In each simulation different waiting times were studied and the total duration of the simulation was $t=3\times 10^4$. In the last part of this work (Sec. III.C), the repulsive barrier is switched off to study the competition between phase separation and the attractive glass transition in the state $\phi_p=0.80$ (50 independent simulations were also performed for this case).

\section{Results and Discussion}

Upon increasing the attraction strength at constant colloid density ($\phi_c=0.40$) the attractive glass transition is found at $\phi_p=0.4265$, as estimated from the analysis of the density correlation functions, self-diffusion coefficients and viscosity according to MCT \cite{puertas03,puertas05,puertas05b}.

As stated above, we study the aging dynamics for two states quenched to within the non-ergodic region, $\phi_p=0.50$ and $\phi_p=0.80$. The former is just within, and the latter well within, the non-ergodic region.  
(Note that the minimum in the potential is $V_{min}=-8 k_BT$ and $V_{min}=-12.8 k_BT$ for $\phi_p=0.50$ and $\phi_p=0.80$, respectively, for the average sized particles.)
The structure and dynamics is tackled for both quenches and compared with an ergodic fluid close to the transition, namely $\phi_p=0.42$. Some theoretical predictions are also tested. The effect of the repulsive barrier is studied in the final subsection.

\begin{figure}
\psfig{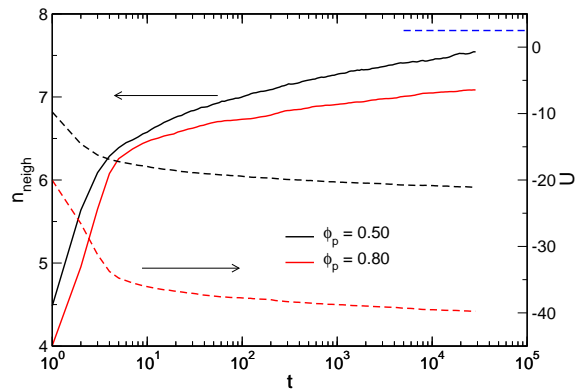}
\caption {\label{energy} Evolution of the mean number of neighbors per particle (continuous lines, left scale) and energy per particle (broken lines, right scale) for both quenches: $\phi_p=0.50$ (black lines) and $\phi_p=0.80$ (red lines).The blue horizontal line in the upper right corner marks the mean number of neighbors for the equilibrium state $\phi_p=0.42$.}
\end{figure}

\subsection{Structure}

The instantaneous quench from $\phi_p=0$ provokes the formation of bonds between particles. Since both states are above the percolation line in density, a bonded cluster extending throughout the system is formed at zero time (using the interaction range, $2\xi=0.2a$, to define the length of a bond). 

In Fig. \ref{energy} the evolution of the mean number of neighbors (bonds) per particle, and of the energy per particle, are presented for both quenches. As the bonds between particles form, the energy per particle decreases. This process slows down as the waiting time increases, but a steady state is not reached in the simulation. Interestingly, more bonds are formed in the quench to the lower $\phi_p$, although the energy per bond is smaller in this case. However, the equilibrium system at $\phi_p=0.42$ reaches a larger number of bonds per particle (horizontal blue line in the upper right corner of the figure), at least in the time studied in the simulation. Note that, in comparison with repulsion driven glasses (with broadly similar potential minima) \cite{kob97}, the energy evolves much more in the attractive case. This can be expected given the relatively low density ($\phi_c=0.40$) of our system. Recall that, at $\phi_p=0$ no bonds are present in the system; but as soon as the quench is performed a highly bonded state exists. This state evolves with time, and can alter considerably since particles are not strongly caged by their hard-core repulsions at this modest $\phi_c$. This leaves much more room for energy drift in attractive glasses than in the repulsion-driven case.

\begin{figure}
\psfig{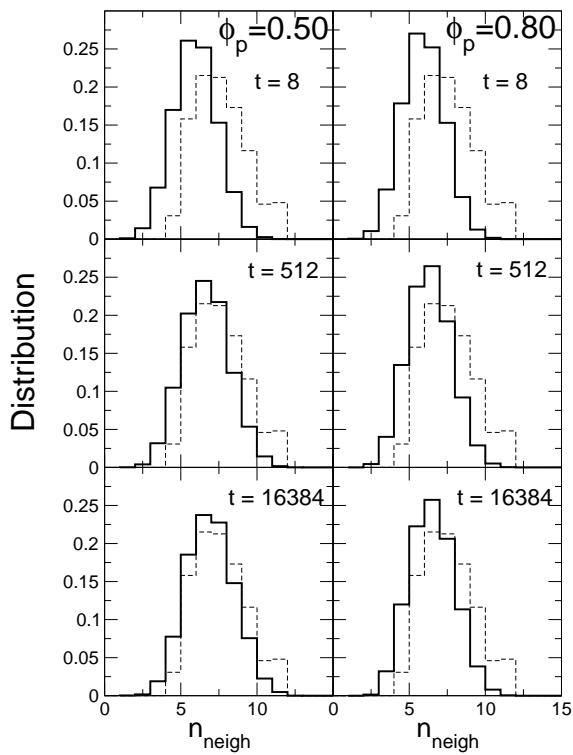}
\caption {\label{nneigh-dist} Distribution of bonds per particle for $\phi_p=0.50$ and $\phi_p=0.80$ at different waiting times, as labeled. The discontinuous line shows the distribution in the equilibrium system at $\phi_p=0.42$.}
\end{figure}

The intricate structure of the network can induce a heterogeneous distribution of the bonds per particle. In Fig. \ref{nneigh-dist} the evolution of the distribution of bonds per particle is presented for both states at different waiting times and compared with the distribution for the equilibrium system at $\phi_p=0.42$. All of the distributions (including that of the system in equilibrium) are wide, due to the heterogeneity of the system, but single-peaked. As aging proceeds, the distribution moves to higher numbers of neighbors per particle causing an increase in the mean number, but never reaches the distribution at equilibrium. Thus, the difference in the mean number of neighbors shown in Fig. \ref{energy} is caused by a displacement of the distribution towards larger number of neighbors, and not by the development of stronger heterogeneities in the structure of the system.

\begin{figure}
\psfig{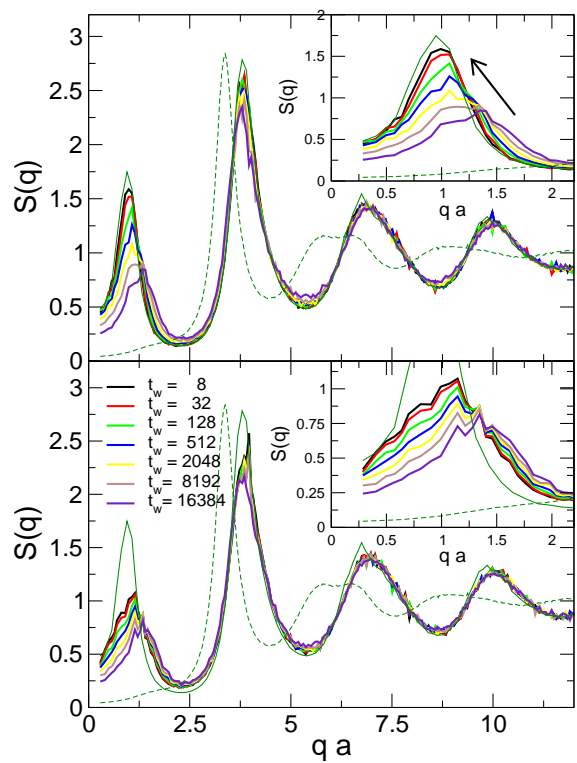}
\caption {\label{sq} Structure factor for different waiting times (as labeled), for $\phi_p=0.50$ (upper panel) and $\phi_p=0.80$ (lower panel). $S(q)$ for $\phi_p=0$ and $\phi_p=0.42$ are also presented (broken and continuous thin green lines, respectively). The insets show the low-$q$ region for every state magnified.}
\end{figure}

The structure factor of the system is presented in Fig. \ref{sq}, for both quenches, at different waiting times. 
Increased local order is noticed in the main peak in $S(q)$; this slightly increases with waiting time, and the oscillations at high-$q$ become stronger. But the most prominent feature in the evolution of $S(q)$ is the appearance of a low-$q$ peak, which grows and moves to smaller wavevectors with waiting time (see insets). Since the system has bond percolation from the outset, this peak is associated not with cluster formation, but with voids appearing and growing (thus the peak moves to smaller $q$) as the bonding state evolves. Accordingly, the structure factor for the homogeneous system at $\phi_p=0$ shows no peak in this region, and the neighbor peak is displaced to larger wavevectors in both quenches (particles come closer to each other).

In agreement with Fig. \ref{energy}, the fluid system at $\phi_p=0.42$ shows a higher neighbor peak and stronger oscillations, since there are more bonds in the equilibrium state than in the quenches (the difference being larger for $\phi_p=0.80$). Also, the low-$q$ peak is highest in the equilibrium state (on the time scales studied in the simulation), and higher for lower $\phi_p$ in the non-ergodic region. 

The overall trend is consistent with slow restructuring of the system to reach a lower energy; for stronger bonds the system is more hindered on its way towards the energy minimum (even though this is deeper in principle). This hindrance is characteristic of aging systems, although in attractive glasses it is caused by bonding, as opposed to core-core repulsions in repulsive glasses.

The low wavevector peak in $S(q)$ resembles a low angle peak found in experiments on colloidal gelation and aggregation \cite{carpineti92,poon95,poulin99,segre01}. The origin of the experimental peak has not fully been determined; since gelation inhibits liquid-gas separation, the peak is probably the signature of frustrated demixing \cite{faraday03,kroy04}. It also obeys the factorization rule typical of spinodal decomposition \cite{furukawa85,poon95,poulin99}. In our case, the low $q$ peak is not caused by an arrested phase separation, but it is an intrinsic characteristic of the system induced by the long range repulsive barrier in the interaction. It is a signature of the trend to microphase separation. Therefore, it will stop growing and moving to smaller $q$ since the system does not demix. Its behavior is thus similar to the experimental one but not of the same origin. We show below that the dynamics of the system is independent of the origin of this peak in our simulations.

\begin{figure}
\psfig{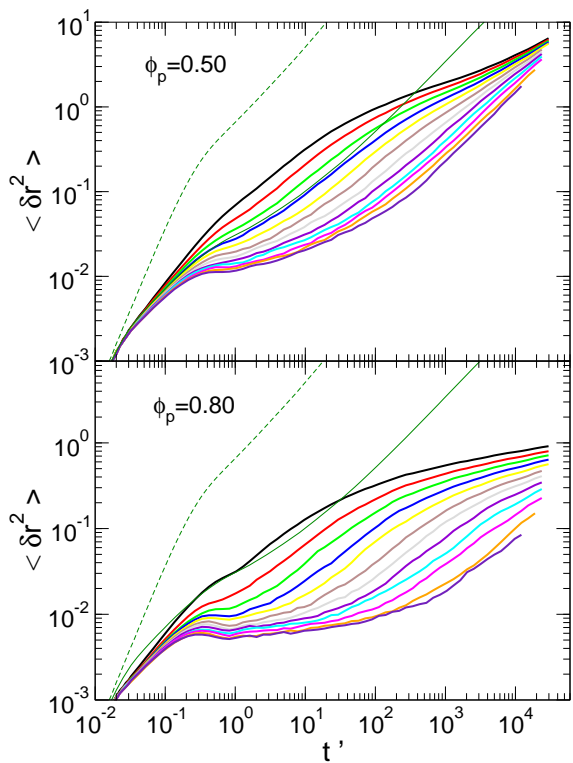}
\caption {\label{msd} Mean squared displacement as a function of $t'=t-t_w$ for both quenches at different waiting times: from left to right $t_w=8$, $16$, $32$, $64$, $128$, $256$, $512$, $1024$, $2048$, $4096$, $8192$ and $16384$. The thin green lines show the mean squared displacements for $\phi_p=0$ (broken line) and $\phi_p=0.42$ (continuous line).}
\end{figure}

\subsection{Dynamics}

The stiffness of the system with increasing attraction strength discussed above is apparent also in its dynamics. In Fig. \ref{msd} the mean squared displacement is presented for different waiting times, as a function of $t'=t-t_w$, for both quenches. As time proceeds, the system is more solid-like, and the displacements of particles is hindered. For comparison, $\langle \delta r^2 (t') \rangle$ for the equilibrium system at $\phi_p=0.42$ is also presented. At $\phi_p=0.80$, even for short waiting times, the dynamics of the out-of-equilibrium systems is slower than the fluid one, although the mean number of bonds per particle is significantly smaller (see Figs. \ref{energy} and \ref{nneigh-dist}). 

The comparison between both quenches shows that the system at $\phi_p=0.80$ slows down much faster than at $\phi_p=0.50$. At large waiting times, the former shows a clear plateau in $\langle \delta r^2 (t') \rangle$ at $\simeq 6\times 10^{-3}a^2$, which extends for three decades in time. On the other hand, the state at $\phi_p=0.50$ only shows a hint of the plateau, which is located at longer distances, and the mean squared displacement reaches larger values.

\begin{figure}
\psfig{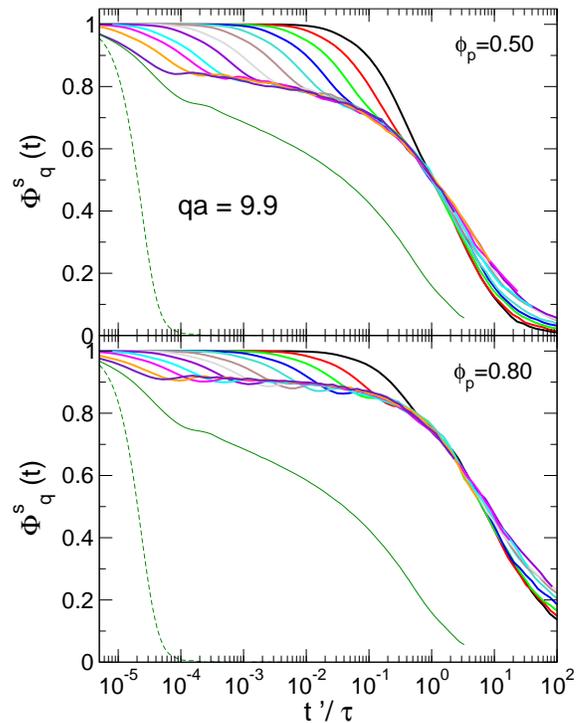}
\caption {\label{scaling} Incoherent density correlation function, $\Phi_q^s(t)$, time rescaled to match the decay from the plateau. Different lines show different waiting times (same as Fig. \ref{msd}), increasing from right to left. The thin green lines show $\Phi_q^s(t')$ for $\phi_p=0$ (broken line) and $\phi_p=0.42$ (continuous line).}
\end{figure}

The self part of the intermediate scattering function, or incoherent density correlation function, shows similar features as the mean squared displacement: a progressive development of a plateau, clearer and higher for $\phi_p=0.80$ than for $\phi_p=0.50$. In Fig. \ref{scaling}, the correlation functions for $qa=9.9$ (third peak in $S(q)$) are presented for the same waiting times as Fig. \ref{msd}, with the time axis rescaled in each case to data-collapse the final decay from the plateau. Very good scaling is observed for the initial decay from the plateau, whereas deviations are observed at long times. The scaling can be significantly improved if only the data with not-too-small waiting times are used, as shown below (Fig. \ref{scaling-nobarrier}).

As expected, the plateau in the correlation function is clearer (flatter) in the case of $\phi_p=0.80$, whereas no real plateau is seen at $\phi_p=0.50$. When compared to $\phi_p=0$ (broken thin line), more than five decades in time of slowing down are observed. It is also noticeable that the shape of the relaxation curve changes with the attraction strength; it is more stretched in the equilibrium system ($\phi_p=0.42$, continuous thin line) than in the aging states. We interpret these findings as arrested dynamics induced by particle bonding, indicated by the existence of a plateau in $\Phi_q$ for $\phi_p=0.80$, even with reversible bonds that are not stabilized artificially \cite{zaccarelli03}. 

\begin{figure}
\psfig{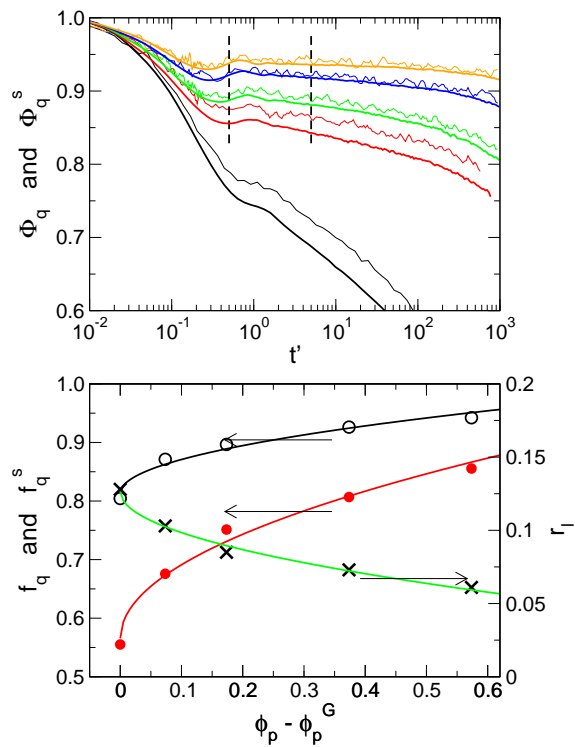}
\caption {\label{short-times} Upper panel: Density correlation functions, coherent (thin lines, $qa=9.6$) and incoherent (thick lines, $qa=15.6$), for different states. From bottom to top: $\phi_p=0.42$ (fluid), $0.50$, $0.60$, $0.80$ and $1.00$. Lower panel: Nonergodicity parameters (coherent, open circles and incoherent, closed circles), and localization lengths vs. the distance to the transition point. The lines are square root fits to the data.}
\end{figure}

The evolution of the plateau can be observed in Fig. \ref{short-times}, where we study the $\beta$ regime for $\phi_p=0.50$, $0.60$, $0.80$ and $1.00$. In this case, correlation functions were measured, for time intervals up to $t'=1000$, for $100$ different waiting times in the range $10^4 < t_w < 6\times 10^4$. (Six independent quenches were made for every state.) No evolution was observed, within the short times probed in Fig. \ref{short-times}, for either $\Phi_q(t)$ or the M.S.D., as expected from Figs. \ref{msd} and \ref{scaling}. The height of the plateau in $\Phi_q(t)$ grows as the state is deeper in the glass, and the nonergodicity parameter is calculated from the value of the function in the $\beta$ regime, i.e. between the two vertical dashed lines in Fig. \ref{short-times}. Similarly, the localization lengths obtained from the plateau in the M.S.D. decreases as $\phi_p$ increases.

\begin{figure}
\psfig{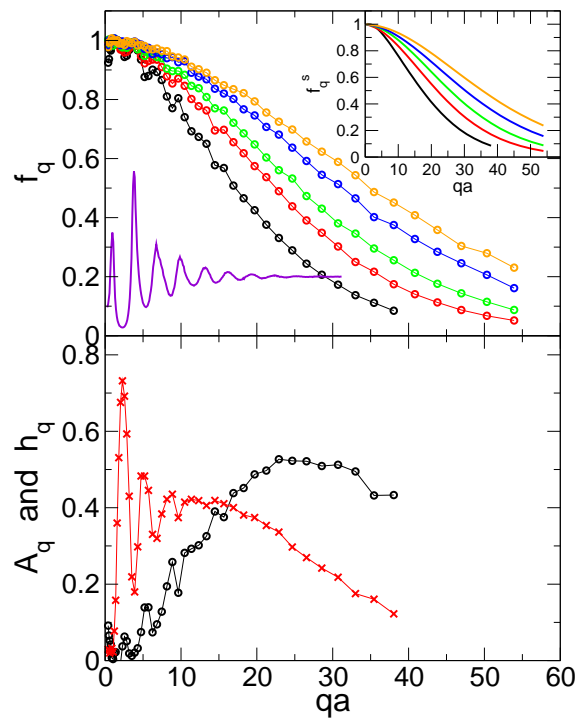}
\caption {\label{fq} Upper panel: Nonergodicity parameter as a function of the wavevector for states progressively deeper in the glass. From top to bottom: critical, $\phi_p=0.50$, $0.60$, $0.80$ and $1.00$. The inset presents the incoherent non-ergodicity parameter for the same states. Lower panel: Amplitude of the square-root fitting to $f_q$, $A_q$ (black circles), and amplitude of the first order in the von Schweidler expansion of the $\alpha$-decay, $h_q$ (red crosses).}
\end{figure}

Both the nonergodicity parameter (left scale in the lower panel, presented for the coherent, $f_q$, and incoherent functions, $f_q^s$) and the localization length, $r_l$, show square root behaviours as a function of the distance to the transition: $f_q=f_q^c+A_q \sqrt{\phi_p-\phi_p^G}$. Here $f_q^c$ is the critical non-ergodicity parameter, $A_q$ is a $q$-dependent amplitude and $\phi_p^G$ is the transition point (with similar expressions holding for $f_q^s$ and $r_l$). As an example, the lower panel of Fig. \ref{short-times} presents $f_q$ for $qa=9.6$, $f_q^s$ for $qa=15.6$ and $r_l$ (right y-scale). Similar trends have been reported with molecular glass formers \cite{gotze99}, and recently with $m$-Toluidine \cite{comez05}, yielding a square root dependence on the temperature shift from the transition point. Interestingly, MCT predicts such a dependence close to the transition on the glass side; but the square root regime observed here  extends much further into the glass than does the asymptotic approximation. (The latter's range of validity can be found by numerical MCT calculations for our system \cite{henrich06}.)

The wavevector dependence of both the self and coherent nonergodicity parameters is presented in Fig. \ref{fq} for the same states as Fig. \ref{short-times}. Whereas $f_q$ shows oscillations in phase with $S(q)$, $f_q^s$ decays monotonically, both of which are in agreement with MCT expectations. For states deeper in the glass, the $q$-distribution of $f_q$ or $f_q^s$ becomes wider, implying the decrease of the localization length. The critical nonergodicity parameter is obtained from the square-root fitting to the data, and agrees, to within our numerical accuracy, with results from equilibrium simulations \cite{puertas05}. The critical $\phi_p$ for the transition has been taken as $\phi_p^G=0.4265$, in accordance with previous results from equilibrium. The amplitude of the square-root in the fitting of $f_q$ vs. $\phi_p^G-\phi_p$ also oscillates in (anti)phase with the structure factor (lower panel of Fig. \ref{fq}). Similar results were obtained with molecular glasses \cite{gotze99,comez05}, {\em i.e.}, $f_q$ oscillates in phase with $S(q)$ and $A_q$ in antiphase; our results extend over the whole range of wavevectors. In ideal MCT, $A_q$ should be equal to the first order amplitude in the von Schweidler expansion of the structural $\alpha$-decay, $h_q$ \cite{gotze99}. The comparison of the two fitted quantities, shown in the lower panel of Fig. \ref{fq}, is not satisfactory. This indicates that the fitted square-root dependence in Fig. \ref{short-times} is not the asymptotic MCT one; this concurs with the fact that the apparent square-root behavior extends over a much wider range of $\phi_p-\phi_p^G$ than does the asymptotic MCT result.

\begin{figure}
\psfig{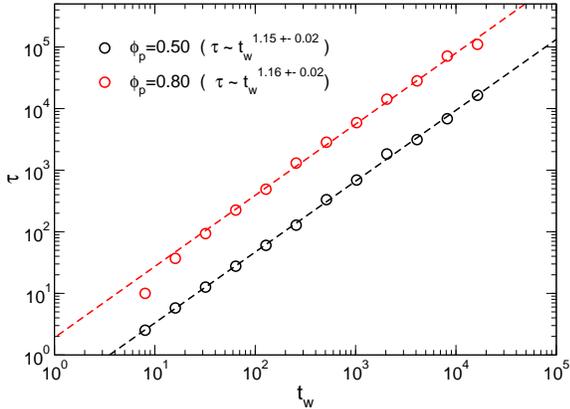}
\caption {\label{tw} Relaxation time normalized with $\tau_{t_w=8}$ as a function of the waiting time, for $\phi_p=0.50$ and $0.80$, and the power-law fittings. The blue line is $\tau \sim t_w$, i.e. simple aging.}
\end{figure}

The relaxation time as a function of the waiting time is presented in Fig. \ref{tw} for both the $\phi_p = 0.50$ and the $\phi_p = 0.80$ quench states. The ratio $\tau_{t_w}/\tau_{t_w=8}$ is given as a measure of the relaxation time; this is the necessary time factor in Fig. \ref{scaling} to collapse all of the curves onto the correlator at $t_w=8$. Power laws are observed, $\tau \sim t_w^{\alpha}$ with exponents bigger than one in both cases. Simple aging, for which $\tau \sim t_w$, is also drawn in the figure; the fits show clearly that $\alpha > 1$, so that the system evolves `faster' (in some sense) than its age. We might expect this behavior to cross over to linear aging eventually; meanwhile it is reminiscent of ``superaging'' observed in spin glasses \cite{kawamura98,zotev03}, although any mechanistic link between those systems and ours is rather unclear.

The data collapse we observe under time rescaling of the correlation functions has also been seen in other structural glasses\cite{kob97}, again accompanied by a power law relation between $\tau$ and $t_w$. In repulsion-driven glasses, this behavior sets in at $t_w=0$ \cite{kob97}, but in colloidal gelation the power law dependence holds only at large waiting times \cite{cipelletti00,abou01,bellour03}. It is preceded by a different behavior, crossing over either from another power-law \cite{darjuzon03}, or from a more ``explosive'' regime (found also in experiments) \cite{cipelletti00,abou01,bellour03}. In these cases, however, the kinetics of aging may be influenced by phase separation,which our study was specifically designed to avoid. 

\begin{figure}
\psfig{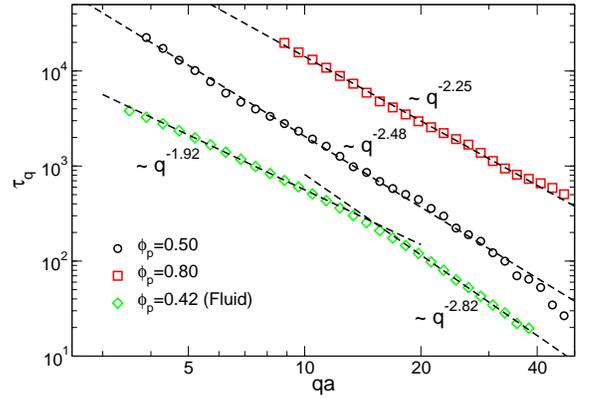}
\caption {\label{tauq} Wavevector dependence of the relaxation time ($\Phi_q^s(\tau_q)=f_q/2$), for $\phi_p=0.50$ (black circles) and $0.80$ (red squares) and $t_w=4096$, and for the equilibrium state at $\phi_p=0.42$ for comparison (green diamonds). The dashed lines are power-law fittings with exponents ($-2.48\pm 0.02$) for $\phi_p=0.50$ and ($2.25\pm 0.02$) for $\phi_p=0.80$, and ($2.82\pm 0.03$) for $\phi_p=0.42$ at high $q$ and ($1.92\pm 0.03$) for low $q$.}
\end{figure}

The wavevector dependence of the time scale, defined via $\Phi_q^s(\tau_q)=f_q/2$, is presented in Fig. \ref{tauq} for both quench states at a waiting time $t_w=4096$ (chosen because it shows a significant part of the structural relaxation, allowing determination of $\tau_q$). In both states, the time scale follows a power law with $q$, with exponents greater than $2$. In fluid states close to the glass transition, according to MCT, this exponent is $-1/b$ where the von Schweidler exponent $b$ describes stretching of the $\alpha$-decay. The results for the equilibrium state at $\phi_p=0.42$ are presented for comparison, showing one exponent $(2.82\pm 0.03)$ at high $q$ (yielding $b=0.36$, compatible with $b=0.37$ obtained by direct von Schweidler fitting \cite{puertas03}), and another exponent $(1.92\pm 0.03)$ for lower $q$, approaching the hydrodynamic limit. The exponents obtained here for our two quenches (around $2.4$) are smaller than that of equilibrium states \cite{puertas03}, yielding higher values of $b$, and hence less stretched decays of the density correlation function. This is indeed observed directly, in Fig. \ref{scaling}. Of the two quenches, the state at $\phi_p=0.50$ gives a higher exponent for the $\tau_q$ behavior, and its correlation function is more stretched indicating that the relation between stretching and wavevector dependence of the time scale apparently holds in the glass.

\begin{figure}
\psfig{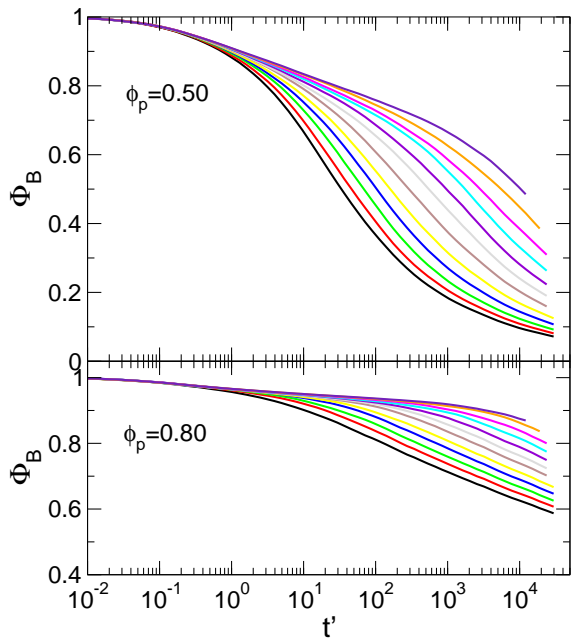}
\caption {\label{bonds} Bond correlation function for the same waiting times as Fig. \ref{msd} from left to right and $\phi_p=0.50$ (upper panel) and $\phi_p=0.80$ (lower panel). Note the different y-scale in both panels.}
\end{figure}

We study next the evolution of the bond correlation function with the waiting time. The bond correlation function, $\Phi_B(t')$, is defined as the fraction of bonds that have existed uninterruptedly since $t'=0$. (Recall $t'=t-t_w$.) This function is presented in Fig. \ref{bonds} for different waiting times, for both $\phi_p=0.50$ and $0.80$. In both cases $\Phi_B$ decays slower as the waiting time increases, indicating that the structure becomes stiffer, in agreement with the evolution of $\Phi_q$ and of the structure factors presented above. It should be noted also that the bonds are much more stable when the quench is made to $\phi_p=0.80$, even at short waiting times, as expected from the preceding results.

The bond correlation function decays in both states, what implies that bonds are never permanent, even at long waiting times (i.e. a plateau in $\Phi_B(t')$ is not observed), although the fraction of bonds that remain unbroken is much larger for $\phi_p=0.80$ than $\phi_p=0.50$. In contrast, the mean squared displacement (Fig. \ref{msd}) and density correlation function (Fig. \ref{scaling}) indeed showed arrested motion of the particles at long waiting times for $\phi_p=0.80$. These results from the bond correlation function thus indicate that, although the relaxation of density fluctuations and the diffusion of particles are both arrested, individual bonds are breaking (and new ones are being formed) within the cage or bonding network around each particle.

\begin{figure}
\psfig{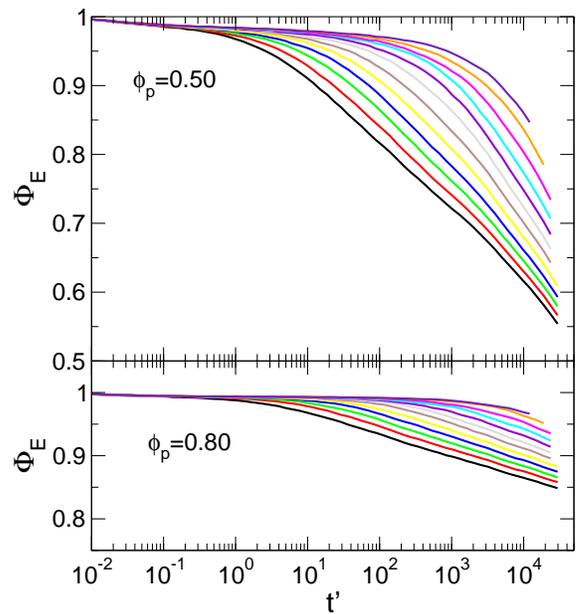}
\caption {\label{cage} Same as Fig. \ref{bonds} for the cage correlation function (see text for the definition). Note that different y-scales are used in the two panels.}
\end{figure}

This is further supported by analyzing the environment, or cage, correlation function, $\Phi_E(t')$, which is defined as the fraction of bonds at time $t'$ that existed at $t'=0$ (regardless of whether they broke in between). The function is presented for the same waiting times, in both quench states, in Fig. \ref{cage}. As required, $\Phi_E(t')>\Phi_B(t')$ since bonds that break and reform are included in the former but not the latter. Note that for the shortest waiting time and $\phi_p=0.50$, less than $10\%$ of the bonds have remained continuously unbroken, but, on average, more than half of a particle's bonds are still part of its environment.

At long waiting times the environment correlation function plateaus at a very high value, which extends to longer times as the waiting time grows. Thus, the network of particles comprising the system becomes stiffer and its relaxation takes longer. In the quench to $\phi_p=0.80$ at the largest $t_w$, $\Phi_E$ has decayed only a few percent, implying that most of the particles have been unable to escape their network of bonds. 

These results for $\Phi_E$ fully support the conclusions drawn from $\Phi_q$; in fact, $\Phi_E$ could almost be viewed as a density correlation function at a wavevector corresponding to the size of the neighborhood of the particles. Interestingly, some bonds can break even at long waiting times (as shown in Fig. \ref{bonds}), although the local structure is frozen. Our results show that the dynamical arrest is a collective phenomenon that can be induced by bonds with finite energy, and thus finite life-time, in contradiction with previous works where it was concluded that the attractive glass was unstable due to the finite bonding energy \cite{zaccarelli03}.

\begin{figure}
\psfig{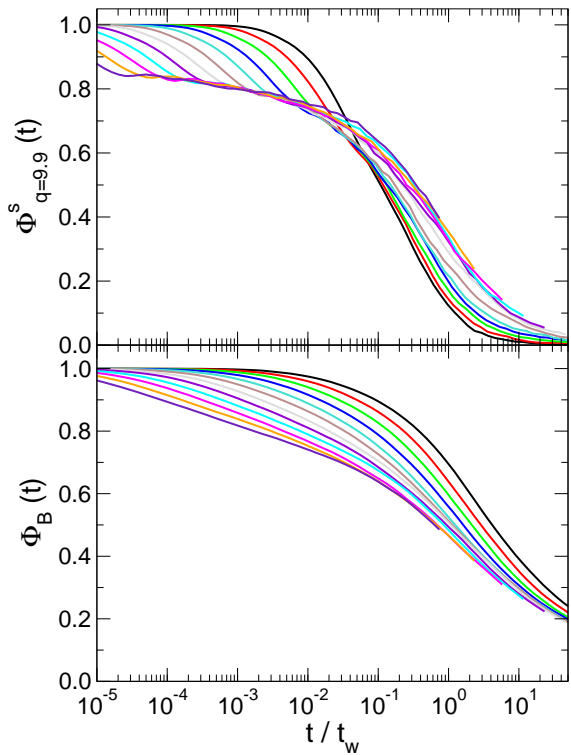}
\caption {\label{scaling-tw} Density correlation function (upper panel) and bond correlation function (lower panel) time rescaled with waiting time, $t_w$, for $\phi_p=0.50$. Different colors correspond to different waiting times, as in Fig. 4.}
\end{figure}

To further exploit the relationship between structural arrest and bond formation, we time-rescale the bond correlation function $\Phi_B$ and density correlation function $\Phi_q$ using the waiting time, $t_w$; see Fig. \ref{scaling-tw} for $\phi_p=0.50$. After the rescaling, neither $\Phi_q$ or $\Phi_B$ collapse, but the density correlation functions cross at $x=t/t_w\approx 10^{-1}$, whereas no crossing is observed for $\Phi_B(x)$. Therefore, at constant $x$ beyond the crossing point, i.e. $x>10^{-1}$, $\Phi_q^s(x)$ is larger for longer waiting times, whereas $\Phi_B(x)$ is smaller. In other words, the time scale for the slowing down of structural relaxation grows faster than the `bond life time'. (Note that a unique bond life time cannot be defined, however, because the decays of the bond correlation functions do not collapse for different waiting times.) This decoupling between structural relaxation and bond formation again points to the collective origin of the attraction-driven glass transition. The implications of Fig. \ref{scaling-tw} differ from previous results for equilibrium systems where the dynamics (as measured by the mean squared displacement) and the bond correlation function were scaled using thermodynamic properties (the second virial coefficient, $B_2$)\cite{foffi05b}. In our case, no quantity can be used collapse by time-rescaling both $\Phi_q$ and $\Phi_B$,  because they evolve differently. 
 Currently we are unable to reconcile our findings with those of Ref. \cite{foffi05b}, if the latter are to apply at low temperatures (high interaction strength). We can only suspect that there is some fundamental difference in either the physical specification of the system or the computational methodology. However, we have not been able to identify such a difference.

It is also worth noticing in Fig. \ref{scaling-tw} that the rescaled curves, both for $\Phi_q$ and $\Phi_B$, approach each other more closely at long waiting times. It is tempting to expect collapse at still longer waiting times, implying a crossover from the ``superaging'' regime reported above to a simple aging one, thus leaving $t_w$ as the only physical time scale in the system. Our present simulations, however, only hint at this possibility, which we can neither prove nor rule out. Similar results are obtained for $\phi_p=0.80$ (data not shown), both in the decoupling of the time scale for structural relaxation and bond life time, and in the suggested simple aging regime for long waiting times.

\begin{figure}
\psfig{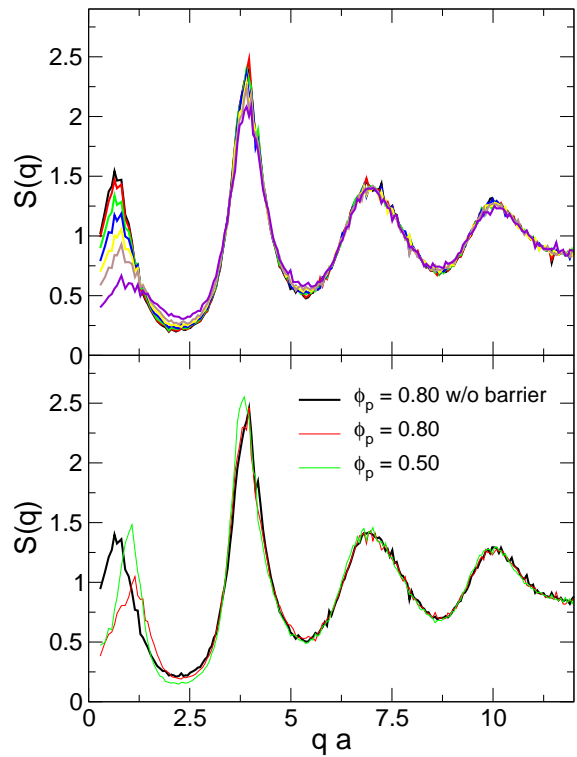}
\caption {\label{sq-nobarrier} Upper panel: Structure factor of the system without repulsive barrier for different waiting times, as labeled. Lower panel: Comparison of the structure factors with the same waiting time, $t_w=4096$, for the states $\phi_p=0.50$, $\phi_p=0.80$, and $\phi_p=0.80$ without the repulsive barrier, as labeled.}
\end{figure}

\subsection{Effect of the repulsive barrier}

In order to test the effect of the repulsive barrier in the interaction potential, introduced to prevent phase separation but possibly inducing microphase separation, we have performed simulations without the barrier. The system, thus, is closer to the experimental colloid-polymer mixtures where gels have been deeply studied \cite{pham04,poon95,segre01}. In previous simulation works, it was shown that the liquid-gas separation is arrested by the attraction driven glass transition, in agreement with experiments \cite{darjuzon03,foffi05,kroy04}. We show below that indeed this is the case in our system and we analyze the effect of the repulsive barrier on the dynamics of the system.

In Fig. \ref{sq-nobarrier} we show the evolution of the structure following the quench. The structure factor develops a low-q peak which moves to lower wavevector and grows, but apparently saturates during the simulation time -- for comparison, in the system with $\phi_p=0.35$ below the glass point $\phi_p^G=0.4265$ the low-q peak moves out below the accessible wavevectors, for $t_w>128$. The structure factors of states with and without the repulsive barrier are compared for the same waiting time in the lower panel of Fig. \ref{sq-nobarrier}. The peak is placed at lower wavevectors and is higher for the system without barrier. These results indicate that the phase separation is prevented by the glass transition, resulting in a heterogeneous system with voids and dense regions. The effect of the repulsive barrier in the potential is then to reduce the size and concentration of voids.

\begin{figure}
\psfig{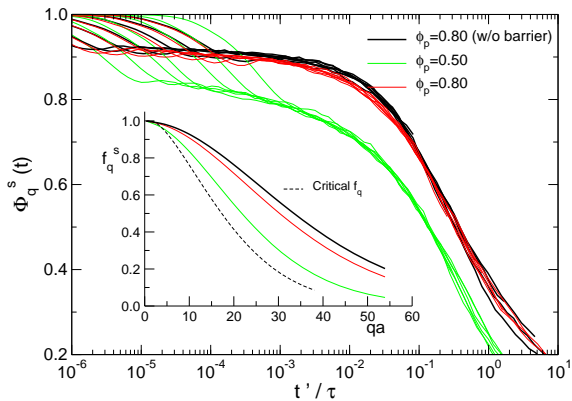}
\caption {\label{scaling-nobarrier} Master decay of the incoherent density correlation function for $\phi_p=0.80$ without repulsive barrier (thick black lines) and with it (thin red lines) and $\phi_p=0.50$ with it (thin green lines). The master decays have been obtained from systems with $t_w=256$, $512$, $1024$, $2048$, $4096$, $8192$, $16384$, from right to left in upper part. The inset shows the non-ergodicity parameters for  the same three states, and the critical one (from Ref. \cite{puertas02}), as a function of the wavevector. }
\end{figure}

An analysis similar to that of Sec. III.B can be conducted to study the dynamics of the arrested system. The mean squared displacement and density correlation function show a slowing down of the dynamics as the waiting time increases, in a two-step process, similar to glass aging. The density correlation functions can be rescaled to overlap the decay from the plateau as shown in Fig. \ref{scaling-nobarrier} for $\phi_p=0.80$ without barrier, where only waiting times above $t_w=256$ are included. For comparison, the master function for $\phi_p=0.80$ and $\phi_p=0.50$ with the repulsive barrier are also included. Note that the scaling of the curves is much better than in Fig. \ref{scaling} since only data with $t_w>256$ is used. The two master functions for $\phi_p=0.80$ are very similar over the whole decay from the plateau, indicating that the long time relaxation of the system is only slightly affected by the repulsive barrier. The time scale of this relaxation as a function of the waiting time (not shown) also obeys a power law, with exponent larger than one: $\tau \sim t_w^{1.20}$, very close to the exponent found in the system with barrier, $1.16$. 

\begin{figure}
\psfig{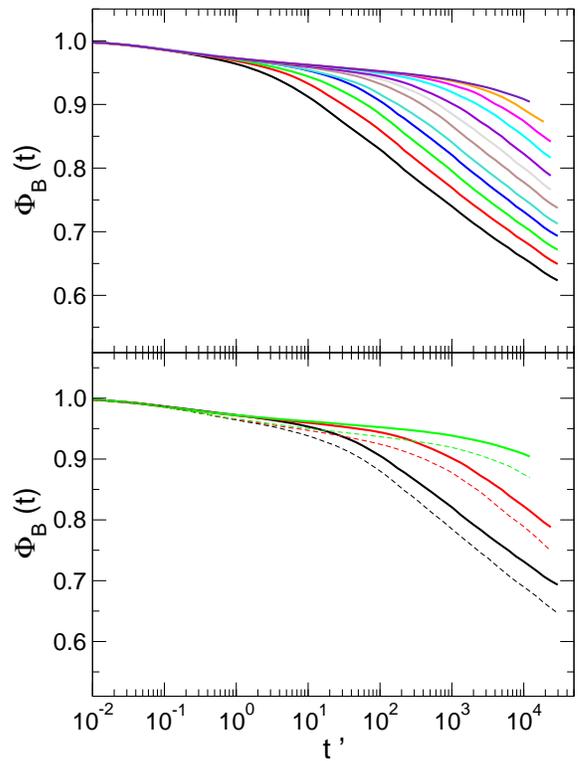}
\caption {\label{bonds-nobarrier} Bond correlation function for the system with $\phi_p=0.80$ without the repulsive barrier, for the same waiting times as Fig. \ref{msd} (upper panel). The lower panel shows a comparison of the bond correlation function for systems at $\phi_p=0.80$ without the barrier (thick lines) and with the barrier (dashed lines) for $t_w=64$, $1024$ ad $16384$ (from bottom to top). }
\end{figure}

The non-ergodicity parameters for the incoherent correlation function are presented in the inset of Fig. \ref{scaling-nobarrier} for the same states. The agreement between the two $f_q^s$ at $\phi_p=0.80$ indicates that the localization length is similar in both cases, and that the origin of the transition is the same: bond formation. Fig. \ref{bonds-nobarrier} presents the evolution of the bond correlation function for the state $\phi_p=0.80$ without the repulsive barrier (upper panel). This function decays even at very long waiting times, although the mean bond-life time grows continuously. The comparison of the bond correlation functions for the system with and without the barrier (lower panel), shows completely similar trends, and that the bonds for the system without the barrier are slightly stronger.

The results presented in this subsection indicate that the repulsive barrier plays a minor role in the behaviour of the system beyond the glass transition, although it has a tremendous effect below it suppressing the liquid-gas separation. The competition between glassy physics and phase transition results in a system with arrested phase separation, whose properties are largely similar to those of the system without the phase transition. This study also rules out the possibility that microphase separation, if it occurs in our system, is primarily responsible for the results presented above.

\section{Conclusions}

In conclusion, we have shown that aging of attraction driven glasses is broadly analogous to the aging of repulsion driven glasses, albeit with some specificities arising from the different driving mechanism in either case: bond formation for the former, steric hindrance for the latter. Specific differences appear, for instance, in the values of localization length and in $f_q$. In attraction driven systems, as in repulsive ones, the mean squared displacement and density correlation functions display a plateau followed by a decay, which moves to longer times as the waiting time grows. $\Phi_q(t')$ can be time-scaled to collapse during the decay from the plateau for different waiting times, and the decay time follows a power-law dependence on the waiting time, with an exponent slightly bigger than unity. The non-ergodicity parameter increases as the state is deeper in the glass, following an apparent square-root behavior, and oscillates in phase with the structure factor, decaying at larger wavevectors with increasing $\phi_p$, implying a shortening localization length. The bond and environment correlation functions show that as the aging proceeds and more bonds are established, a stiffer network of particles forms whose relaxation time increases; but that the bonds themselves are not permanent. However, whereas the time scale for structural relaxation grows faster than the waiting time, the bond life time increases slower than $t_w$. This decoupling of structural relaxation from single bond breaking illustrates the collective character of the dynamical arrest. Yet, from our simulations, it cannot be ruled out that for large enough $t_w$ both time scales approach simple aging behavior, scaling directly with $t_w$. The effect of the repulsive barrier (introduced to suppress liquid-gas 
separation) on the dynamics is quite weak beyond the glass transition; this suggests that the transition is primarily induced by bonding, and that aging in attractive glasses is dominated by the evolution of the bonds rather than by any slow evolution of the phase separation pattern.

Important differences can be observed between the quenches to $\phi_p=0.50$ and $\phi_p=0.80$. The former is slightly above the transition line, and does not show clear plateaus in $\Phi_q$, the mean squared displacement or $\Phi_E$. The stretching of the correlation functions is also different (more stretched closer to the transition point). Nevertheless, scaling of the density correlation functions was fulfilled, and the exponent of the relaxation time as a function of the waiting time is similar for both states. From the structural point of view, the glass with lower attraction strength allows for more restructuring: the holes are bigger, as noticed by a higher low-$q$ peak in the structure factor, and more bonds per particle are formed. Interestingly, our molecular dynamics results strongly indicate that bonds of finite energy, which are therefore reversible on the time scale of the simulation, can induce dynamical arrest at high enough interaction strength. These conclusions are different from those drawn from previous works \cite{zaccarelli03} although the results themselves are not very different from ours. The decoupling of structural relaxation from single bond breaking represents another aspect where our results differ from previous findings \cite{foffi05b}, if those can be extrapolated to low temperatures.

The softness of the glass at low attraction strength (close to the boundary, $\phi_p=0.50$) may arise from a non-vanishing population of highly mobile particles detected in the fluid phase close to the transition point \cite{puertas05,puertas04}. This population coexists with a rigid network of particles, and reduces as the attraction strength is increased, being completely absent at $\phi_p=0.80$ \cite{puertas06}. The origin of the softness can thus perhaps be traced back to the strong dynamical heterogeneities in the fluid.

Finally, in addition to the results presented in this paper, we have found recently that many dynamical quantities for the fluid states close to the glass line can be accounted for within MCT, not only in its universal qualitative predictions, but also quantitatively \cite{puertas05,henrich06}. Furthermore, the dynamic heterogeneities of the glass decrease as the state point moves deeper into the attractive glass \cite{puertas06}, in agreement with findings for repulsive glasses \cite{vollmayr02,vollmayr05}. All of these results support the argument that attraction driven glasses share many universal properties with repulsion driven glasses, albeit with some differences that are attributable to the different driving mechanism. 

\begin{center}
{\sc Acknowledgments}
\end{center}

A.M.P. acknowledges the financial support by the Spanish Ministerio de Educaci\'on y Ciencia (under project MAT2003-03051-CO3-01). Work in the UK funded by EPSRC Grant GR/S10377. M.F. and A.M.P. acknowledge support by the DAAD (ref. D/04/39929) and MEC-AI (ref. HA2004-0022).

\end{document}